\documentclass[12pt]{iopart}

\usepackage[utf8]{inputenc}
\usepackage[OT4]{fontenc}

\def\beq{\begin{equation}}
\def\eeq{\end{equation}}
\def\bea{\begin{eqnarray}}
\def\eea{\end{eqnarray}}
\def\beax{\begin{eqnarray*}}
\def\eeax{\end{eqnarray*}}

\def\ket#1{{\left|#1\right>}}

\def\matel#1#2#3{{\langle#1|#2|#3\rangle}}

\def\mat#1#2#3#4{\left(\begin{array}{cc}#1&#2\\#3&#4\end{array}\right)}
\def\vect#1#2{\left(\begin{array}{c}#1\\#2\end{array}\right)}

\begin{document}
\title{Generator Coordinate Method framework for Double Beta Decay}
\author{Andrzej Bobyk$^a$, Wiesław A. Kamiński$^b$}
\address{${}^{a)}$Faculty of Mathematics, Computer Science and Landscape Architecture, The John Paul II Catholic University of Lublin, ul.~Konstantynów 1H, 20-708 Lublin}
\address{${}^{b)}$Department of Informatics, Maria Curie-Skłodowska University in Lublin, ul.~Akademicka 9, 20-033 Lublin, Poland}

\begin{abstract}
We propose a consistent prescription for the derivation of the particle number and angular momentum projected QRPA (PQRPA) equation in the Generator Coordinate Method (GCM) framework for calculation of NME's of double-beta decay of axially deformed nuclei. We derive closed formulae for the calculation of excitation energies and wave functions of the intermediate nucleus.
\end{abstract}
\vspace{2pc}
\submitto{\PS}

\section{Introduction}

Neutrinoless double beta decay $0\nu\beta\beta$, if observed experimentally, would provide unique information on the neutrino properties \cite{bib1, bib2, bib3}. The $0\nu\beta\beta$ decay process and the associated nuclear matrix elements (NME), necessary to extract these properties from experimental data, were investigated up to now using several approaches, including the quasiparticle random phase approximation (QRPA) \cite{bib1}, the interacting shell model \cite{bib4, bib5}, the interacting boson model \cite{bib6, bib7} and the projected Hartree-Fock Bogoliubov model \cite{bib9}. Although the Generator Coordinate Method (GCM) has been applied previously in the context of double beta decay \cite{bib8, bib81}, it has been used solely over PHFB calculations to account for configuration mixing effects. In the present work we propose the GCM framework to deliver consistent number- and angular momentum projected QRPA (PQRPA) approach for calculation of NME's of double-beta decay of axially deformed nuclei.

\section{Projection method}

Nuclear states with well-defined angular momentum quantum numbers in the laboratory frame can be constructed by projecting the corresponding components of the nuclear states, defined in the intrinsic frame of the nucleus \cite{peierls}:
\beq
	P^J_{MK} = \frac{2J+1}{8\pi^2}\int d\Omega \, {D^J_{MK}(\Omega)}^*\, R(\Omega)
\label{PJKM},
\eeq
where $R(\Omega) = e^{-i\alpha J_3}\, e^{-i\beta J_2}\, e^{-i\gamma J_3}$ is the rotation operator and $\Omega$ represents Euler angles $\alpha$, $\beta$ and $\gamma$ ($d\Omega = d\alpha\, d\gamma\, \sin\beta\, d\beta$).

In the case of axial symmetry of the nucleus, this operator reduces to the form: \cite{hara}:
\beq
	P^J_{MK} = \left(J+\mbox{$\frac{1}{2}$}\right)\int_0^\pi d\beta\, \sin\beta\, d^J_{MK}(\beta)\, R(\beta),
\label{PJKM_axial}
\eeq
with $R(\beta) = e^{-i\beta J_2}$, where $d^J_{MK}(\beta)$ are the so-called "small" (reduced) real-valued Wigner matrices. The particle number projector for protons and neutrons can be expressed in the similar form:
\beq
	P_{Z,N} = \frac{1}{(2\pi)^2} \int_0^{2\pi} \int_0^{2\pi} d\zeta\, d\eta\, e^{i(\zeta Z_0 + \eta N_0)}\, 
	R_Z(\zeta)\, R_N(\eta),
\eeq
where $R_Z(\zeta) = e^{-i\zeta Z}$ and $R_N(\eta) = e^{-i\eta N}$. From now on we will use the shorthand notation $R= R(\beta)\, R_Z(\zeta)\, R_N(\eta)$ and $P= P^J_{MK}\, P_{Z,N}$.

To obtain the excited states in the intermediate nucleus we need to construct the charge-changing operator, that creates a pair of quasi-nucleons. Because for a deformed nucleus with axial symmetry, the conserved quantum numbers are the $z$-component of the angular momentum projection on the symmetry axis of the nucleus ($K$) and the spatial parity ($\pi$), this operator can be expressed in the form:
\beq
	B^\dagger_{(pn)K^\pi} = \left[ \beta^\dagger_p \beta^\dagger_n \right]_{K^\pi},
\eeq
with $m_p + m_n = K, (-1)^{l_p+l_n} = \pi$, where $\beta^\dagger_p$ and $\beta^\dagger_n$ are, respectively, the quasi-proton and the quasi-neutron creation operators in the deformed basis. The corresponding annihilation operators appear in the projected HFB equation \cite{Zeh}:
\beq
	\matel{\Phi_0}{\beta_{k'}\beta_k (H-E^P_0)P_0}{\Phi_0}
\eeq
for the ground-state $\ket{\Phi_0}$ and the (projected) ground-state energy, given by:
\beq
	E^P_0 = N^{-1}_0\matel{\Phi_0}{HP_0}{\Phi_0},
\eeq
with the norm $N_0 = \matel{\Phi_0}{P_0}{\Phi_0}$. Please note, that $P_0$ in the above expressions is the projected operator on the corresponding quantum numbers for the ground state of the initial/final nuclei and not the excited states in the intermediate nucleus. In our approach we not only project out the non-physical components in the ground states of the initial and final nuclei but in the intermediate states as well.

To construct these states within the PQRPA framework we use the derivation proposed in \cite{fed}, where they are obtained as trial functions in the form, given by the GCM ansatz:
\beq
	\ket{\Psi} = \int dz\, f(z)\, P\ket{\Phi(z^*)},
	\label{GCM}
\eeq
where
\beq
	\ket{\Phi(z)} = P\, \exp \left(\sum_{pn} z_{pn}\, \beta^\dagger_p \beta^\dagger_n \right) \ket{\Phi_0}.
\eeq

This leads to the well-known integral Hill-Wheeler equation \cite{rs}, that using the so-called gaussian overlap approximation (GOA) can be cast in the differential form. Within this approximation the norm overlap can be expressed as:
\beq
	N(z, z'^*) = N_0 \exp \left\{ 
		(s^*,\, s) \vect{z'^*}{z} + \frac{1}{2} (z,\, z'^*) \mat{M}{T}{T^*}{M^*} \vect{z'^*}{z}
	\right\}
\eeq
with the coefficients $s_\kappa$, $M_{\kappa\lambda}$ and $T_{\kappa\lambda}$ given by:
\bea
	s_\kappa & = & \frac{\partial}{\partial z_\kappa} \ln \left. N \right|_{z=z'=0}
		= N_0^{-1}\matel{\Phi_0}{\beta_n \beta_p P}{\Phi_0} \\
	M_{\kappa\lambda} & = & \frac{\partial^2}{\partial z_\kappa \partial z'^*_\lambda} \ln \left. N \right|_{z=z'=0}
		= N_0^{-1}\matel{\Phi_0}{\beta_n \beta_p P \beta^\dagger_{p'}\beta^\dagger_{n'}}{\Phi_0} - s_\kappa s^*_\lambda \\
	T_{\kappa\lambda} & = & \frac{\partial^2}{\partial z_\kappa \partial z_\lambda} \ln \left. N \right|_{z=z'=0}
		= N_0^{-1}\matel{\Phi_0}{\beta_n \beta_p \beta_{n'}\beta_{p'}P}{\Phi_0} - s_\kappa s_\lambda,
\eea
where $\kappa = \{p, n\}$ and $\lambda = \{p', n'\}$ are combination of proton/neutron indices, fulfilling condition $m_p + m_n = K, (-1)^{l_p+l_n} = \pi$.

Following \cite{hara} to calculate the contractions, we can now obtain explicit expressions for these coefficients, {\em i.e.}:
\bea
	\label{int1}
	s_\kappa & = & \frac{J+\frac{1}{2}}{(2\pi)^2 N_0}
	\int_0^\pi \int_0^{2\pi} \int_0^{2\pi} d\omega\, \rho(\omega) \sqrt{\det U(\omega)}\, \e^{-i(\zeta\Omega_Z+\eta\Omega_N)}
		B_{np}(\omega) \\
	\label{int2} 
	M_{\kappa\lambda} & = & \frac{J+\frac{1}{2}}{(2\pi)^2 N_0}
	\int_0^\pi \int_0^{2\pi} \int_0^{2\pi} d\omega\, \rho(\omega) \sqrt{\det U(\omega)}\, \e^{-i(\zeta\Omega_Z+\eta\Omega_N)} \\
		& & \times\big(B_{np}(\omega) A_{p'n'}(\omega) - C_{np'}(\omega) C_{pn'}(\omega) + C_{nn'}(\omega) C_{pp'}(\omega)\big) 
		\nonumber \\
		& & - s_\kappa s^*_\lambda 
		\nonumber \\
	\label{int3}
	T_{\kappa\lambda} & = & \frac{J+\frac{1}{2}}{(2\pi)^2 N_0}
	\int_0^\pi \int_0^{2\pi} \int_0^{2\pi} d\omega\, \rho(\omega) \sqrt{\det U(\omega)}\, \e^{-i(\zeta\Omega_Z+\eta\Omega_N)} \\
		& & \times\big(B_{np}(\omega) B_{n'p'}(\omega) - B_{nn'}(\omega) B_{pp'}(\omega) + B_{np'}(\omega) C_{pn'}(\omega)\big) 
		\nonumber \\
		& & - s_\kappa s_\lambda 
		\nonumber
\eea
with $\omega = \{\beta, \zeta, \eta\}$ and $\rho(\omega) = \sin\beta\, d^J_{MK}(\beta)\, e^{i(\zeta Z_0 + \eta N_0)}$ and the $A, B, C$ and $U$ matrices given by:
\bea
	U_{\nu\nu'}(\omega) & = & (u_\nu u_{\nu'} \e^{i\varphi} + v_\nu v_{\nu'} \e^{-i\varphi}) W_{\nu\nu'}(\beta) \\
	V_{\nu\nu'}(\omega) & = & (u_\nu v_{\nu'} \e^{i\varphi} - v_\nu u_{\nu'} \e^{-i\varphi}) W_{\nu\overline{\nu'}}(\beta)
\eea
\beq
	C(\omega) = U^{-1}(\omega), \quad A(\omega) =  V^*(\omega) C(\omega), \quad B(\omega) = C(\omega) V(\omega)
\eeq
In the above expressions, $\Omega_Z$ ($\Omega_N$) is the total number of proton (neutron) levels and $\varphi=\zeta$ 
($\varphi=\eta$) for protons (neutrons). This means that these matrices are block diagonal, e.g.:
\beq
	U = \mat{U^{\rm (p)}}{0}{0}{U^{\rm (n)}}, \quad 
	\det U(\omega) = \det U^{\rm (p)}(\beta, \zeta) \cdot \det U^{\rm (n)}(\beta, \eta).
\eeq

It should be noted, that the integrals (\ref{int1})--(\ref{int3}) can be evaluated numerically exactly with a relatively small number of integration points, using formulae given in \cite{hara} and \cite{chatt}. Moreover, since the intermediate nucleus is an odd-odd nucleus it is more convenient in practice (albeit equivalent) to project not onto $Z$ and $N$, but onto $A/2=(N+Z)/2$ and $T_3=(N-Z)/2$, that are both integers.

In the above expressions $u$ and $v$ are the HFB transformation coefficients and $W_{\nu\nu'}(\beta)$ are the matrix elements of the $R(\beta)$ operator between the single-particle states in the deformed basis. They can be calculated as follows:

Matrix elements of the $J_2$ component of the (intrinsic frame) angular momentum in the axially-symmetric deformed basis $\ket{am}$ ($a$ denotes all the other quantum numbers except $m$) can by found by expanding the states of this basis on the eigenstates $\ket{nljm}$ of the spherical harmonic oscillator. We get:
\bea
	\matel{a'm'}{J_3}{am} & = & \delta_{aa'} \delta_{mm'}\cdot m \\
	\matel{a'm'}{J^2}{am} & = & \delta_{mm'} \cdot \sum_{nlj} C^a_{nlj} (C^{a'}_{nlj})^* \cdot j(j+1),
\eea
where $C^a_{nlj}$ expansion coefficients of the states $\ket{am}$ on the spherical states:
\beq
	\ket{am} = \sum_{nlj} C^a_{nlj} \ket{nljm}.
\eeq

We define in the usual way "ladder" operators of the angular momentum:
\beq
	J_{\pm} = J_1 \pm i\, J_2.
	\label{jpm}
\eeq
Because $J_+ = J_-^\dagger$, we get:
\bea
	J_+ \ket{am} & = & \lambda_{am}\, \ket{a, m+1} \\
	J_- \ket{a, m+1} & = & (\lambda_{am})^*\, \ket{am}
\eea
One can easily show that $J_- J_+ = J^2 - J_3^2 - J_3$, thus, keeping Condon and Shortley phase convention ($\lambda_{am} > 0$) we get:
\beq
	\lambda_{am} = \sqrt{\left\langle j(j+1) \right\rangle_a - m(m+1)},
\eeq
where  $\left\langle j(j+1) \right\rangle_a = \sum_{nlj} \left|C^a_{nlj}\right|^2 \cdot j(j+1)$ is the expectation value of the $J_2$ operator in the $a$ shell. Resolving (\ref{jpm}) we obtain matrix elements of $J_2$:
\beq
	\matel{a'm'}{J_2}{am} = \frac{i}{2} \delta_{aa'} \left(\delta_{m', m-1} - \delta_{m', m+1}\right)
	\sqrt{\left\langle j(j+1) \right\rangle_a - mm'}.
\eeq

Matrix elements of the rotation operator $R(\beta) = e^{-i\beta J_2} = e^{r(\beta)}$ can be found, using corresponding algorithms for calculation of exponent of real antisymmetric matrices \cite{gallier}, since the matrix of the $r(\beta)$ operator is block-wise antisymmetric:
\beq
	r_{am,a'm'}(\beta) = \frac{\beta}{2}\delta_{aa'} \left(\delta_{m', m-1} - \delta_{m', m+1}\right) r_{amm'},
\eeq
where $r_{amm'} = \frac{\beta}{2}\sqrt{\left\langle j(j+1) \right\rangle_a - m m'}$. Let primed indices enumerate columns of the ${\bf r} = \hbox{diag}({\bf r}_a)$ matrix and let quantum numbers $m$ run values from $-m_{\rm max}$ up to $m_{\rm max}$ for each block $N_a \times N_a$. Blocks ${\bf r}_a$ take then the form:
\beq
	{\bf r}_a = \left( \begin{array}{ccccc}
		0         & -r_{amm'} & 0         & \cdots    & \cdots \\
		r_{amm'}  & 0         & -r_{amm'} & 0         & \cdots \\
		0         & r_{amm'}  & 0         & -r_{amm'} & \cdots \\
		\vdots    & 0         & r_{amm'}  & 0         & \cdots  \\
		\vdots    & \vdots    & \vdots    & \vdots    & \ddots
	\end{array}\right),
\eeq 
that allows us easily express ${\bf R} = e^{\bf r}$ as $\hbox{diag}(e^{{\bf r}_a})$.

\section{Derivation of the PQRPA equations}

For the derivation of projected QRPA (PQRPA) equations we follow the procedure, described by Federschmidt and Ring \cite{fed}, with several hints on the resolution of the integral equation and determination of the wave function (\ref{GCM}), cast by Jancovici and Schiff \cite{janc}. In the spirit of the GOA, if we stop after second order in $z, z'^*$ in the approximation of the ratio of the Hamiltonian and the norm kernels:
\beq
	h(z, z'^*) = \frac{H(z, z'^*)}{N(z, z'^*)},
\eeq
we obtain an equation, that resembles already the well-known QRPA form:
\beq
	h(z, z'^*) = E^P_0 + \frac{1}{2}(z, z'^*)\mat{A}{B}{B^*}{A^*}\vect{z'^*}{z}
\eeq
with the generalized PQRPA matrices:
\bea
	A_{\kappa\lambda} & = & N_0^{-1}\matel{\Phi_0}{\beta_n \beta_p (H-E^P_0) P \beta^\dagger_{p'}\beta^\dagger_{n'}}{\Phi_0} \\
		& = & N_0^{-1}\matel{\Phi_0}{\beta_n \beta_p H P \beta^\dagger_{p'}\beta^\dagger_{n'}}{\Phi_0} 
		- E^P_0 (M_{\kappa\lambda} + s_\kappa s^*_\lambda) \nonumber \\
	B_{\kappa\lambda} & = & N_0^{-1}\matel{\Phi_0}{\beta_n \beta_p \beta_{n'}\beta_{p'} (H-E^P_0) P}{\Phi_0} \\
		& = & N_0^{-1}\matel{\Phi_0}{\beta_n \beta_p \beta_{n'}\beta_{p'} HP}{\Phi_0}
		- E^P_0 (T_{\kappa\lambda} + s_\kappa s_\lambda), \nonumber 
\eea
that can be calculated using the recursive formula for the antisymmetrized sum of products of contractions, of the type given by Hara and Iwasaki \cite{hara}, analogous to the ordinary Wick theorem. Actually, we use this relation to avoid explicit writing of $(8-1)!! = 105$ terms and simply treat this recursion numerically.

The hermitian matrix $M$ can be then decomposed into $WW^\dagger$ and boson operators can be introduced:
\beq
	\label{bosons}
	B_\kappa^\dagger = \sum_\lambda W_{\lambda\kappa} z_\lambda, \quad 
	B_\kappa = \sum_\lambda (W^{-1})_{\lambda\kappa} \frac{\partial}{\partial z_\lambda},
\eeq
which allows the Hill-Wheeler equation \cite{rs} to be transformed into differential equation. Defining the PQRPA phonon operators:
\beq
	A_\mu^\dagger = \sum_\kappa (X_{\kappa\mu} B_\kappa^\dagger - Y_{\kappa\mu} B_\kappa)
\eeq
we end up with the projected QRPA (PQRPA) equation ($\Omega_\mu$ is the phonon energy):
\beq
	\mat{A}{B}{B^*}{A^*}\vect{X}{Y}_\mu = \Omega_\mu \mat{M}{0}{0}{-M^*}\vect{X}{Y}_\mu
\eeq
for the forward- and backward-going amplitudes $X$ and $Y$, respectively. One should note, that without the projection, this equation reduces to the usual QRPA equation, since then coefficients $s_\kappa$ and $T_{\kappa\lambda}$ vanish and $M_{\kappa\lambda}$ becomes an unity matrix.

In practice, one more step has to be done, because due to overcompletness of the basis, the coefficients $X$ and $Y$ do not fulfill usual orthogonality relations and removal of zero-energy (spurious) excitation modes is necessary\footnote{Similar problem arises when one concerns the so-called second-order QRPA or extended QRPA with inclusion of scattering terms.}. This step is done along with the decomposition of the matrix $M=WW^\dagger$ by using only those components with non-vanishing eigenvalue, which makes the (effective) $W$ matrix rectangular in principle and reduces the dimension of the PQRPA equation. Finally, the (orthogonalized) boson operators read:
\beq
	C_\mu^\dagger = \sum_\kappa (W^\dagger X)_{\kappa\mu} B_\kappa^\dagger - (W^T Y)_{\kappa\mu} B_\kappa)
\eeq
and the redefined PQRPA matrices gain the form:
\beq
	\tilde{A} = W^\dagger A W, \quad \tilde{B} = W^T B W^*
\eeq
giving the usual shape of the PQRPA equation:
\beq
	\mat{\tilde{A}}{\tilde{B}}{-\tilde{B}^*}{-\tilde{A}^*}\vect{\tilde{X}}{\tilde{Y}}_\mu = 
	\Omega_\mu \vect{\tilde{X}}{\tilde{Y}}_\mu,
\eeq
where $\tilde{X} = W^\dagger X$ and $\tilde{Y} = W^T Y$.

\section{Determination of the wave function}

Te determine the wave function (\ref{GCM}) of the excited intermediate nucleus, we first introduce the transformed weight function:
\beq
	g(z) = \int dz'\, \exp\left(\sum_\mu z_\mu z'^*_\mu\right) f(z')
\eeq
that allows us to transform the Hill-Wheeler equation into a differential equation for $g(z)$ \cite{janc}. Since, from (\ref{bosons}) we see, that $B_\mu$ should annihilate the phonon vacuum, the function $G(z)$, which corresponds to the ground-state solution should fulfill the condition
\beq
	B_\mu G(z) = 0
\eeq
that give us a solution of the form:
\beq
	G(z) = \exp\left(-\frac{1}{2}\sum_{\kappa\lambda} Z_{\kappa\lambda} z_\kappa z_\lambda\right),
\eeq
where $Z_{\kappa\lambda}$ is the solution of the system of linear equations:
\beq
	\sum_\mu X_{\kappa\mu} Z_{\mu\lambda} = Y_{\kappa\lambda}.
\eeq
The excited states can then be constructed by applying the phonon creation operators $B_\mu^\dagger$ to the ground-state $G$ function.

However, because $f(z)$ is a function of real and imaginary parts of $z$ separately, it cannot be determined uniquely. It comes from the fact, that states $\ket{\Phi(z)}$ form an overcomplete set. This is actually of an advantage, because we can use the freedom of choice of $f(z)$ to put some convenient condition on it. Indeed, if we use the identity:
\beq
	g(z) = \frac{1}{\pi^N} \int dz' \exp\left(\sum_\mu z_\mu z'^*_\mu - \sum_\mu |z'_\mu|^2 \right) g(z'),
\eeq
where $N$ is the dimension of the PQRPA problem, a possible solution for $f(z$) reads:
\beq
	f(z) = \frac{1}{\pi^N} \exp\left(-\sum_\mu |z'_\mu|^2 \right) g(z).
\eeq
Finally, the excited state wave function (after replacement of each $z_\kappa$ by $\beta^\dagger_p \beta^\dagger_n$) takes the form:
\beq
	\ket{\Psi} = g(\beta^\dagger_p \beta^\dagger_n)\ket{\Phi_0}.
\eeq

\section*{Acknowledgments} This work was supported by the Polish National Science Centre under the decision number DEC-2011/01/B/ST2/05932.

\section*{References}
\bibliography{art_PhScr}

\end{document}